# Middle-obstacle approach of mapping phase-field model onto its sharp interface counterpart

Chuanqi Zhu*, Yusuke Seguchi, Masayuki Okugawa, Yuichiro Koizumi

*Graduate School of Engineering, Osaka University, 2-1, Yamadaoka, Suita, Osaka, Japan*

September 6, 2023

A new diffuse interface model has been proposed in this study for simulating binary alloy solidification under universal cooling conditions, involving both equilibrium and non-equilibrium solute partitioning. Starting from the Gibbs-Thomson equation, which is the classical theory that describes the dynamics of a sharp interface, the phase-field equation is derived using a traveling wave solution that represents a diffuse interface. To tackle the spurious effects caused by the variation of liquid concentration within the diffuse interface with artificial width, a middle obstacle is introduced to "sharpen" the diffuse interface and an invariant liquid concentration can be found for determining a constant undercooling in the interface normal direction. For slow solidification under equilibrium conditions, the convergence performance of the dendrite tip shows superior invulnerability to the width effect of the diffuse interface. For rapid solidification under non-equilibrium conditions, the output partition coefficients obtained from the steady-state concentration profiles agree with the input velocity-dependent function. The proposed model is promising to be an indispensable tool for the development of advanced alloy materials through the microstructure control of solidification under a wide range of cooling conditions.

## I. INTRODUCTION

In alloy solidification, solute partitioning can be either equilibrium or non-equilibrium, depending on the interface velocity [1, 2]. Local equilibrium is usually assumed within the interface moving at a slow or moderate speed and the solute with limited solubility is partially rejected from the newly-formed solid. The concentrations on the two sides of the steady-state interface lie on the solidus and liquidus of the equilibrium phase diagram. When the interface velocity becomes high, the amount of captured solute exceeds the solubility and the liquid concentration ahead of the interface deviates from the equilibrium liquidus. Consequently, the morphology and concentration distribution in the microstructures resulting from slow and fast solidification processes are distinct. Thus, it is vital to fully describe the process of non-equilibrium solute partitioning, which is dependent on interface velocity, for optimizing the processing conditions during rapid solidification and developing advanced alloy materials.

The continuous growth (CG) model [3, 4] and the local non-equilibrium (LN) model [5, 6] are two established models for describing the non-equilibrium solute partition. The interface is assumed to be a sharp plane in these analytical models. Correspondingly, numerical models including diffuse interfaces were developed in accordance with the analytical ones. They are the so-called parabolic [7–9] and hyperbolic [10] phase-field models, respectively. Despite the physical correctness of these phase-field models, they can hardly be implemented in two-dimensional space to be relevant to the characteristic length of the microstructure because the width of the diffuse interface in these models needs to have a physical size on the nano-scale. If the interface width is artificially enlarged to reduce the computational cost, the numerical models are invalid in matching the analytical ones. It is required to develop numerical schemes that can improve the computational efficiency of the diffuse interface models without losing the connection to the analytical sharp interface models.

The so-called quantitative phase-field model [11–14] is the parabolic model with an artificially wide interface. They can be mapped onto the analytical sharp interface model through thin-interface analysis and additional flux incorporated into the diffusion equation. The simulation results of alloy solidification can possibly be independent of the interface width once a thin interface limit is reached. It should be noted that the convergence of the results from these models is mostly limited to low-speed interfaces under equilibrium conditions. It is still challenging to extend them to simulate high-speed interfaces under non-equilibrium conditions.

* Email address: chuanqizhu1991@outlook.com

Recently, attempts have been made to extend the quantitative scheme for simulating solidification under non-equilibrium conditions in accordance with the CG analytical models. In the work of Tatu et al. [15] and Kavousi et al. [16], the anti-trapping current is modified to regulate non-equilibrium solute partitioning. Results weakly dependent on the interface width can be obtained. Ji et al. [17] reproduced the banded structure consistent with experimental observation. In their model, the solute transport in the diffuse interface region is enhanced by introducing a diffusivity interpolated by a quadratic function, which should be adapted according to the interface width. The model proposed by Steinbach et al. [18] incorporates a third kinetic equation for controlling the exchange of solute atoms between solid and liquid phases within the interface. The results tend to be dependent on the interface width and the kinetic parameter (permeability) needs to be determined by fitting to the experimental or analytical data [19].

In contrast to the variational approach [20], in which the governing equations for concentration and phase field are derived from the variational derivative of the free energy functional, the present work chooses a non-variational way to map the diffuse interface model unto its sharp interface counterpart. Rather than mapping the parabolic or hyperbolic phase-field models unto their analytical counterparts, the velocity-dependent partition coefficients obtained from CG and LN models can be straightforwardly incorporated into the proposed diffuse interface model. The simulations of dendrite growth under equilibrium conditions and excessive solute trapping under non-equilibrium conditions will be demonstrated to verify the applicability of the proposed model.

## II. METHOD

**2.1 Phase-field equation**

The kinetics of the sharp interface is described by the Gibbs-Thomson equation,

$$v = \mu(\Delta T - \Gamma\kappa). \tag{1}$$

The motion of the interface is driven by the contributions from interface undercooling $\Delta T$ and the surface pressure related to curvature $\kappa$ and Gibbs-Thomson coefficient $\Gamma$, which is the ratio of surface energy to fusion entropy. The proportional relationship between the net contribution and interface velocity $v$ is suggested by the kinetic coefficient $\mu$.

In the diffuse interface model, a continuous field parameter is used to represent the bulk phases as well as the interface. The so-called phase field remains constant in the bulk regions and varies smoothly within the interface region. Conventionally, the value of the phase field changes from 1 in the solid to 0 in the liquid, showing a diffuse profile at the interface. For pure substances, the equation that governs the kinetics of the diffuse interface can be expressed by referring to the Gibbs-Thomson equation. The resulting phase-field equation [21, 22] is

$$\frac{1}{|\nabla\phi|}\frac{\partial\phi}{\partial t} = \mu\left[\Delta T - \Gamma\nabla\left(-\frac{\nabla\phi}{|\nabla\phi|}\right)\right], \tag{2}$$

in which interface velocity $v$ and curvature $\kappa$ are replaced by the gradient and local change of the phase field $\phi$. For isothermal conditions when $\Delta T$ is constant throughout the domain, Eqs. (1) and (2) are identical if a proper solution for the phase field with a diffuse interface profile can be found. Compared to the Gibbs-Thomson equation, the phase-field equation facilitates the numerical simulation of pattern formation because the position, curvature, and morphology of the interface can be implicitly and easily obtained from the spatially continuous phase field.

The solution of the phase field can be found by minimizing the total free energy of the whole domain with bulk phases and diffuse interface. Following the derivation in [23], the solution of the steady-state diffuse interface in 1D is specified as,

$$\phi = \begin{cases} 1 & x < -\eta/2 + vt \\ \frac{1}{2} - \frac{1}{2}\sin\frac{\pi}{\eta}(x - vt) & -\eta/2 + vt \leq x \leq \eta/2 + vt \\ 0 & x > \eta/2 + vt \end{cases} \qquad (3)$$

which is the so-called traveling wave solution. The interface width is $\eta$ and the velocity is $v$. The values of the phase field in the solid and liquid regions are 1 and 0, respectively. For position within the interface ($-\eta/2 + vt \leq x \leq \eta/2 + vt$), the solution shows a diffuse contour constructed by the sinusoidal function, which has singularities at the boundaries between the interface and bulk regions. Despite this, the traveling wave solution has its advantage in numerical implementation and is related to the double obstacle function used in the free energy functional. With this solution, Eq. (2) can be reformulated to be,

$$\frac{\partial \phi}{\partial t} = \mu \left\{ \Gamma \left[ \nabla^2 \phi + \frac{\pi^2}{\eta^2}(\phi - \frac{1}{2}) \right] - \frac{\pi\sqrt{\phi(1-\phi)}}{\eta} \Delta T \right\} \qquad (4)$$

## 2.2 Diffusion equation

For a moving interface of a dilute binary alloy with a negative liquidus slope, the solute atoms are rejected from the growing solid into the liquid. The concentration on the solid side is lower than that of the liquid side. Thus, the concentration profile shows a jump at the interface and a diffusion layer in the bulk liquid. This jump can be expressed as

$$c_{iS} < c_{iL}, \qquad (5)$$

in which the superscript $i$ denotes the position at the interface. In the diffuse interface model, this jump is manifested by a continuous transition across the interface (Fig.1). By assuming the interface is a mixture of the solid and liquid phases, the concentrations of solid and liquid are individual field variables and the overall concentration is weighted by the local phase fractions, which is indicated by the local phase field. The relation of the three distinct concentration fields is expressed by,

$$c = c_S \phi + c_L(1 - \phi). \qquad (6)$$

Similarly, the change of the local overall concentration is the sum of changes in individual solid and liquid phases. The fluxes according to Fick's law in solid and liquid phases are also weighted by the local phase factions,

$$J_S = -\phi D_S \nabla c_S, \qquad (7)$$

$$J_L = -(1 - \phi) D_L \nabla c_L, \qquad (8)$$

in which $D_S$ and $D_L$ are diffusivities in solid and liquid phases. The diffusion equation [21, 22] can then be written by complying with the law of mass conservation,

$$\frac{\partial c}{\partial t} = -\nabla (J_S + J_L), \qquad (9)$$

which can be expanded by the chain rule of derivative,

$$\begin{aligned}\frac{\partial c}{\partial t} &= \nabla \cdot \left[D_S \nabla c_S \phi + D_L c_L (1-\phi)\right] \\ &= D_S(\nabla \phi \nabla c_S + \phi \nabla^2 c_S) + D_L\left[-\nabla \phi \nabla c_L + (1-\phi)\nabla^2 c_L\right]\end{aligned} \quad (10)$$

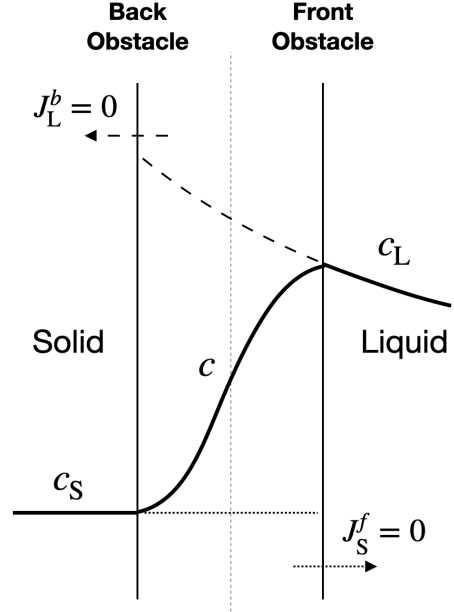

FIG.1 Schematic image of the liquid, solid, and overall concentration fields of a moving interface in the diffuse interface model. The change of overall concentration comes from the sum of changes in the solid and liquid phases, which are implemented separately in computation through the diffusion equation. The fluxes at the limits (back and front obstacles) of liquid and solid phases are set to zero.

Special attention needs to be paid to the boundary conditions of the liquid and solid concentrations at their limits within the interface. During the numerical implementation of the diffusion equation, the fluxes in the liquid and solid should be blocked where the phase fractions become zero:

$$J_L^b = J_S^f = 0. \quad (11)$$

Here, we designate these places as obstacles for solute transport. The limit of the liquid phase is called the back obstacle, while the one of the solid phase is called the front obstacle. The physical meaning of these obstacles is that the solute in the liquid phase cannot be transported to the solid phase by diffusion, and vice versa.

**2.3 Coupling the phase and concentration fields**

In the sharp interface model, the propagation of the interface is accompanied by the inter-diffusion of solute between solid and liquid within the interface as well as the long-range transport of solute away from the interface toward the two sides of bulk phases. The former results in solute partitioning (rejection), while the latter lowers the concentration at the interface. The steady state is realized when the two processes are balanced. Correspondingly, the same processes happen in the diffuse interface model. In the following paragraphs, the details of mapping the diffuse interface model unto its sharp interface counterpart will be explained.

*2.3.1 Concentration changes within interface*

Fig.2 demonstrates the coupled phase and concentration fields in two consecutive steps. The local change of phase field is denoted by $\delta\phi$, which is the amount of liquid that transforms into solid. This change is accompanied by the change of local concentration $\delta c$. If a steady state is realized, the two concentration profiles should keep steady except for the shifted position. To determine the amount of $\delta c$, the diffuse interface is divided into several representative volumes (RVs) to show the specific processes in each individual numerical grid. In Fig.3, the RV is a rectangle with two separate areas representing the mixture of solid and liquid phases with the denoted fractions and concentrations. As the local phase field changes, fractions and concentrations in the current RV on the left-hand side are reconfigured in the next RV on the right-hand side. The fraction of the newly formed solid is $\delta\phi$. Theoretically, for the dilute alloy with a negative liquidus slope, the newly formed solid can not capture all the solute of its previous liquid, which has a concentration of $c_L$. A capture coefficient is then defined to quantify the percentage of the trapped solute in the newly formed solid. Consequently, the amount of rejected solute can be determined to be,

$$\delta c = (1 - \lambda)c_L \delta\phi \tag{12}$$

which is the local descending amplitude of the concentration profile in Fig.2.

The physical ground of solute capturing lies in the process of solute redistribution for reaching the equality of chemical potential. However, because the interface is in motion, the extent of solute redistribution may depend on the velocity of the interface. For most cases when interface velocity is low and moderate, the time is sufficient for the exchange of solute between the local solid and liquid. The capture coefficient $\lambda$ is smaller than one. If the interface velocity is very large, the time for phase transformation and solute redistribution is so limited that the capture coefficient $\lambda$ may approximate unity and no solute is rejected from the solid.

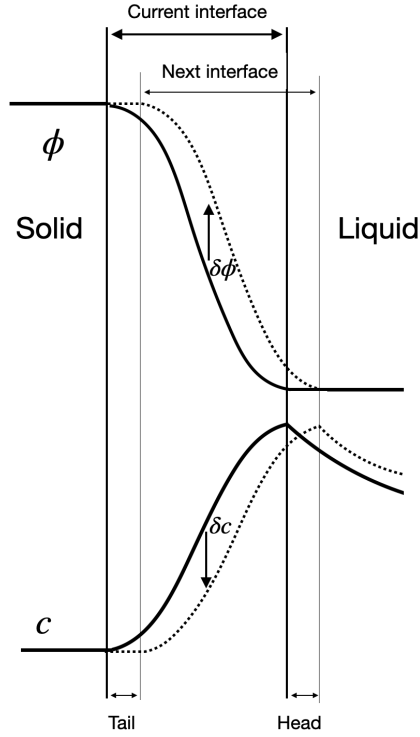

FIG.2 Schematic image of the coupled phase and concentration fields in two sequential steps. As the interface moves forward, the concentration profile shifts in the same direction as a result of solute partitioning and long-range transport.

The capture coefficient is unknown yet. Intuitively, it has a relation to the partition coefficient as both of them approach unity at high interface velocity. The latter is the result of the equality of chemical potential and is related to the inter-diffusion between the solid and liquid phases. The capture process represents the solute flow from the liquid to the solid. A relation between these two coefficients can be found by considering the mass conservation and equality of chemical potential. In the present work, for the sake of simplicity at the beginning of model development, it has been assumed that the capture coefficient equals to the partition coefficient and that all rejected solute atoms are released to the local liquid. This assumption brings about two consequences: one is that the present model is a one-sided model, in which the diffusion in the solid has been neglected; another one is that the capture coefficient can be a function of the interface velocity, which can be obtained from the analytical CG or LN models. Thus, the function of the CG model can be straightforwardly input into the phase field model through the capture coefficient. It can be expressed as

$$\lambda = k(v) = \frac{k_e + v/v_D}{1 + v/v_D}. \tag{13}$$

In future work, the exact relation between the capture and partition coefficients needs to be derived to incorporate the effect of solute diffusion in the solid phase. Also, the partition coefficient function of the LN model will be used to more fully describe the non-equilibrium solute trapping using the proposed phase-field model.

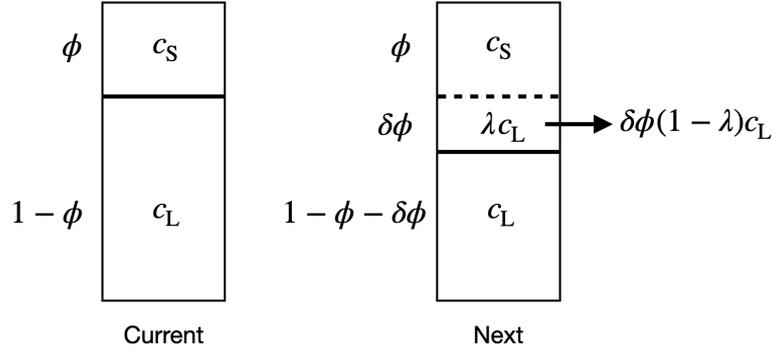

FIG.3 The processes of phase transformation and solute redistribution in a representative volume (RV) of the diffuse interface. For simplicity, the capture coefficient is assumed to be the partition coefficient, which can be the function obtained from either CG or LN model.

*2.3.2 Collect-and-cast operation*

One may ask how to deal with the rejected solute $\delta c$ associated with the change $\delta \phi$ in each RV. In the sharp interface model, the rejected solute should be transported by long-range diffusion into both liquid and solid phases. In the diffuse interface model, it should be released to the local solid and liquid phases. However, problems should arise when the interface width is artificially enlarged. It takes a long distance to transport the solute out of the interface region and the solute concentration in the interface depends on the interface width. In addition, the rejected solute in the tail area (Fig.2) can not be released into the liquid because there is no liquid in that area after the phase transformation. As a result, they should be trapped in the solid. The solute trapping in the diffuse interface model inherently exists in all alloy phase-field models. If the interface width has a physical size on the nano-scale, the parabolic model resembles the sharp interface CG model in modeling solute trapping [7, 8]. However, in the model with an enlarged diffuse interface, the trapping becomes excessive and undesired. An operation on the rejected solute is needed to enhance the solute transport and eliminate this spurious effect caused by the artificially wide diffuse interface.

In the present work, the anti-trapping operation has been accomplished by collecting the rejected solute atoms in each RV and then casting them in the normal direction of the interface. This can be realized by iteratively searching the next grid in the normal direction of the interface during the numerical implementation (See appendix). This anti-trapping operation on the rejected solute $\delta c$ facilitates the solute redistribution within the artificially wide diffuse interface. It is crucial and should be implemented with care to avoid spurious effects that may easily arise. For example, if all the rejected solute atoms are collected and cast into the bulk liquid just in front of the interface, the solute transport is excessively enhanced by this operation in the normal direction of the interface. This enhancement is proportion to the interface width and affects the simulation results. Therefore, a natural and sophisticated way is required to emulate the solute redistribution process in the sharp interface model.

A middle plane of the interface is used to "sharpen" the diffuse interface model. As illustrated in Fig.4, the diffuse interface is divided into the front and back parts by the middle plane, which is called middle obstacle (MO) here. As the dashed arrow in Fig.4 suggests, the rejected solute atoms in the back part should be collected and cast just in front of the middle obstacle, whereas the ones in the front part should be transferred to the local liquid and transported away by diffusion. As a result, the diffuse interface, which has been assumed to be a mixture of solid and liquid, can then be regarded as the extensions from the bulk solid and liquid phases separated by the imaginary "sharp" interface. The tangential diffusion along the arc direction in the diffuse interface turns out to be the diffusion process in the bulk phases.

*2.3.3 Undercooling at the interface*

As the rejected solute atoms $\delta c$ in the two parts separated by the middle obstacle are transported in distinct manners, the profiles of the liquid concentration on the two sides should be quite different. In the front part of the diffuse interface, the profile joins with the middle obstacle and extends its gradient to the diffusion layer in the bulk liquid; in the back part of the diffuse interface, because the liquid concentration is not affected by the collect-and-cast operation, the concentration inside the back part should be quickly flattened by the diffusion process. The liquid concentration just behind the middle obstacle can be regarded as the liquid concentration at the sharp interface. Because it might be independent of the interface width, it is called invariant concentration $c_L^{inv}$, which can be used to determine the interface undercooling.

$$\Delta T = T_m - T_i + c_L^{inv} m_L \tag{14}$$

in which $T_m$, $T_i$, and $m_L$ are the melting point, interface temperature, and liquidus slope. The concentration of each grid within the interface should be determined by searching the nearest invariant concentration in the normal direction through the numerical technique applied in the collect-and-cast operation (See appendix).

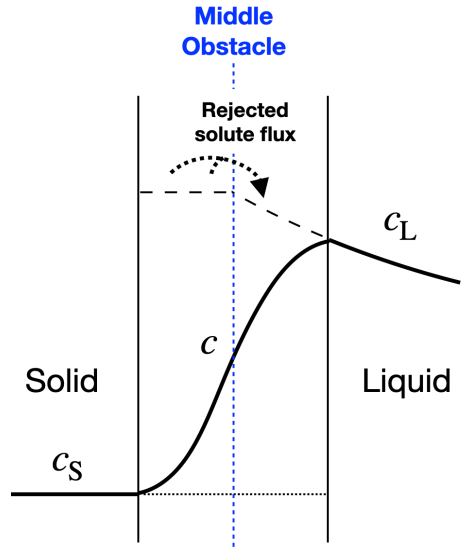

FIG.4 Schematic concentration profiles as a result of the anti-trapping operation and the middle obstacle. The rejected solute atoms in the back part are collected and cast (see appendix) just ahead of the middle obstacle and the ones in the front part are transferred locally to the liquid and transported out of the interface by diffusion. An invariant concentration can be found just behind the middle obstacle.

## III. RESULTS & DISCUSSIONS

The dilute binary alloy system of Al-1.3at.%(3wt.%)Cu is chosen for the simulation of solidification under both equilibirum and non-equilibrium coniditions. The physical parameters [14, 24] are as follows: melting point of pure aluminum, 931 K; diffusivity of copper in the liquid, $2.4 \times 10^{-9}$ m$^2$/s; liquidus slope, -600 K/at.; equilibrium partition coefficient, 0.14; Gibbs-Thomson coefficient, $2.4 \times 10^{-7}$ K · m. Other physical and numerical parameters will be mentioned together with the following simulation results.

## 3.1 Equilibrium conditions

The isothermal temperature is set to be 917K for low-rate solidification under equilibrium conditions. In 1D simulation (Fig.5), the diffuse interface is resolved by 8 grids. The total solute is conserved in the whole domain, suggesting the anti-trapping (collect-and-cast) operation does not change the nominal concentration. As expected, the liquid concentration profile shows a discontinuity at the position of the middle obstacle. In the 2D simulation (Fig.6 inset), the dendrite grows in a moving frame until its tip velocity reaches a steady state. The computational domain has a physical size of $32\ \mu m \times 64\ \mu m$. The resolution of the domain depends on the interface width, which should be adjusted by comparing it with the tip velocity and the resulting diffusion length. Three interface widths (3.0, 1.5, and 0.75 $\mu m$) are used and each of them is resolved by 6 girds. To maintain numerical stability, the kinetic coefficient is chosen to be $0.5 \times 10^{-3}$ m/(sK), which is 200 times smaller that the physical value of 0.1 m/(sK) for atomically-rough surface [26].

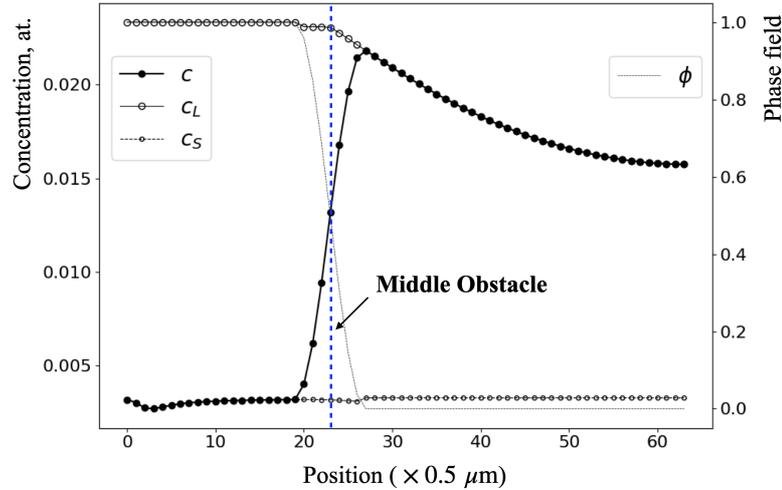

FIG.5 1D simulation result with concentration and phase field profiles. The concentration in the liquid phase shows a discontinuity at the middle obstacle, where $\phi = 0.5$. The liquid phase has the equilibrium concentration in the solid. The same is for the solid phase.

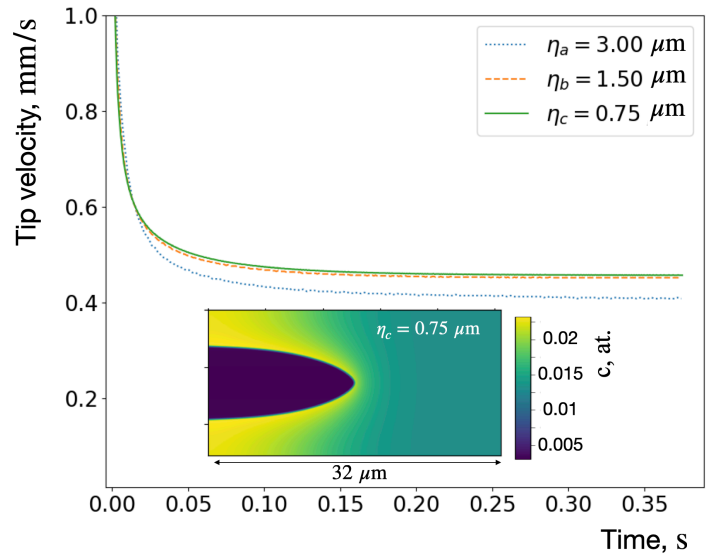

FIG.6 Change of dendrite tip velocity with time under isothermal condition (917K). The inset figure is the concentration map that shows the steady-state dendrite arm in a moving frame at 0.35s.

The convergence of the tip velocity is of great importance in the dendrite growth simulation because it indicates the dependence of the tip dynamics on the artificial interface width. If the tip growth converges with the interface width on the micrometer scale, the tip shape and velocity remain the same even as the interface width is reduced to the nanometer scale. Thus it can then be ensured that the numerical simulation produces the dendrite with the "true" shape, which can hardly be obtained from theoretical analysis [25]. As plotted in Fig.6, three cases of dendrite simulation reach steady states after the physical time of about 0.15s. As the interface width decreases, the tip velocity converges at a value of around 500 $\mu$m/s. The diffusion length $l_D$ [24] ahead of the tip is the ratio of diffusivity to the tip velocity and has a value of about 4.8 $\mu$m. In principle, $l_D$ should be larger than the interface width $\eta$ in phase-field simulation. The ratio of $l_D$ to $\eta$ is a common measure of numerical performance. The smaller the ratio, the higher the efficiency. In the present simulation, the convergence happens with the interface width $\eta_b = 1.5$ $\mu$m considering the difference between the tip velocities in simulations with $\eta_b$ and $\eta_c$ is less than 1.5%. The ratio $l_D/\eta_b$ is about 3.2, which suggests an excellent performance compared to the established quantitative model in the literature [11, 13, 14]. Even with the interface width $\eta_a$, the difference in tip velocity from the converged ones is around 10%. This suggests that the tip velocity is highly invulnerable to the width effect of the diffuse interface owing to the middle obstacle approach. The model is efficient in simulating solidification under equilibrium conditions. The above results suggest that the proposed model has achieved the goal of quantitative and efficient simulation of dendrite growth, which is ubiquitously observed in low-speed alloy solidification.

### 3.2 Non-equilibrium conditions

As the isothermal temperature becomes low and the interface velocity increases, the solute partitioning process turns to be non-equilibrium and dependent on the interface velocity. As the interface velocity reaches as high as about 1 m/s, the diffusion length is shortened to about a few nanometers. Thus, the width of the interface in simulation should be also on the nano-scale. The kinetic coefficient is chosen to have a physical value of 0.1 m/(sK) in the following simulation.

Fig.7 shows the concentration profiles of two high-speed moving interfaces. The diffusion length varies according to the interface velocity. The solid concentrations are equal to the far field liquid concentration, exhibiting the complete trapping phenomenon, which is the typical feature of the steady state during non-equilibrium solidification [18]. The liquid concentration profiles overlap with overall concentrations in the bulk liquid and show a step in the interface, where the invariant concentration lies. The large difference between the invariant and equilibrium concentration suggests the large deviation from the equilibrium and high undercooling at the interface. The undercooling is calculated by referring to the vertical distance to the liquidus in the equilibrium phase diagram. It should be noted that the velocity-dependent liquidus slope [25] and the effect of solute drag [4] have not been considered in the present work.

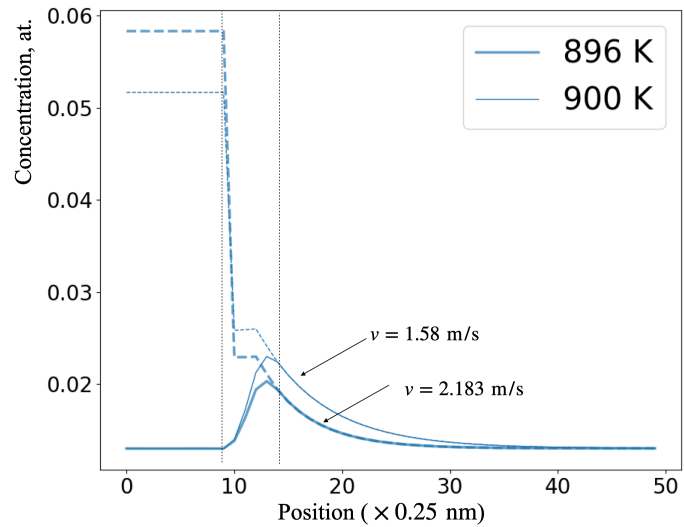

FIG.7 Steady states of rapid growth under isothermal conditions. The complete trapping phenomenon is the typical feature of the steady state of rapid solidification. Interface undercooling is calculated by referring to the liquidus in the equilibrium phase diagram (velocity-dependent liquidus slope and solute drag have not been considered yet).

The partition coefficients obtained from concentration profiles under a range of temperatures are plotted against the interface velocities. The partition coefficient is defined to be the ratio of the invariant concentration to the solid concentration near the interface. When the temperature is high, the complete trapping steady state may not happen, but the instant velocity and partition coefficient can still be measured. These point data from simulations are plotted in Fig.8 along with the function of the CG model, which has been input into the PF model according to Eq.(13). A good agreement between them can be discerned. As shown in Fig.9, when the interface width in the PF model increases, the steady-state velocity and partition coefficient deviate from the results of the small interface width. For the interface width of 5 nm, the deviation (<10%) is obvious but still acceptable and the points stay near the curve of the CG model.

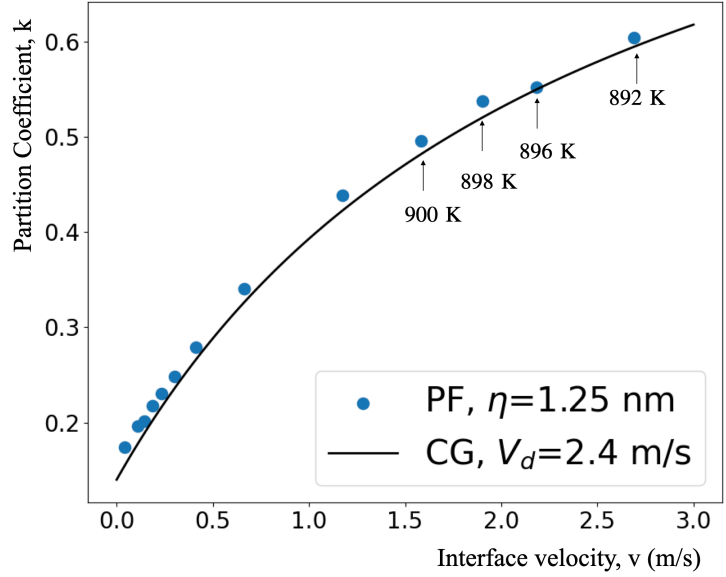

FIG.8 Partition coefficients from continuous growth (CG) model and phase-field (PF) simulation. The diffuse interface width in the PF model is 1.25 nm.

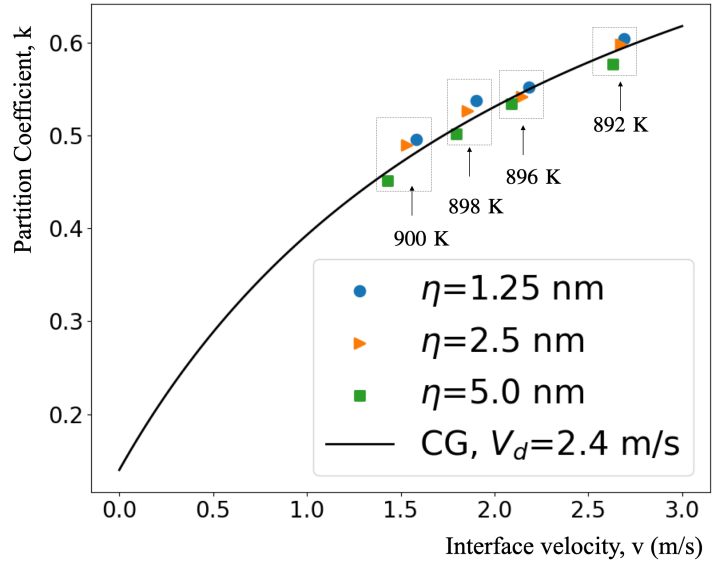

FIG.9 Partition coefficients obtained from PF simulation with different interface widths are compared. All interfaces are in complete trapping steady states. Interface velocity appears to deviate as the interface width increases.

As the PF model is capable of simulating velocity-dependent solute partitioning, it is expected to reproduce experimentally relevant phenomena. For example, the banded structure [27] formed during rapid solidification features in the cyclic concentration distribution with segregated and non-segregated areas, corresponding to equilibrium and non-equilibrium solute partitioning. In the simulation for reproducing the cyclic growth in rapid solidification, a temperature gradient ($10^7$ K/m) is imposed on the domain and the constant cooling rate ($6 \times 10^6$ K/s) gives rise to the pulling velocity of 0.6 m/s. The initial temperature at the interface is 905 K. As shown in Fig.10a, c, the interface grows at a low velocity in the beginning and the concentration jump at the interface suggests diffusion-controlled growth under the equilibrium condition. The interface velocity does not adapt to the pulling velocity immediately because of the accumulating solute ahead of the interface. It only increases gradually along with the decreasing interface temperature. Until a critical state is reached at ~4.3 $\mu$s, the interface velocity surges to be over 3 m/s and the growth is no longer controlled by the solute diffusion but by the atomic attachment (Fig.10b). Because of the high propagation speed, the interface moves far ahead along the temperature gradient, the decreased undercooling relaxes the motion of the interface. The interface slows down and returns to be controlled by diffusion. As Fig.10c shows, the cycle of fast and slow growth modes continues with a steady amplitude of oscillation, which can be discerned from 10 to 16 $\mu$s. Though the resulting concentration distribution (Fig.11) is in one dimension, it is remarkably similar to the banded structure observed in literature (Fig.8 in [27] and Fig.5b in [28] ). The simulation results in 2D will be disclosed in future publication.

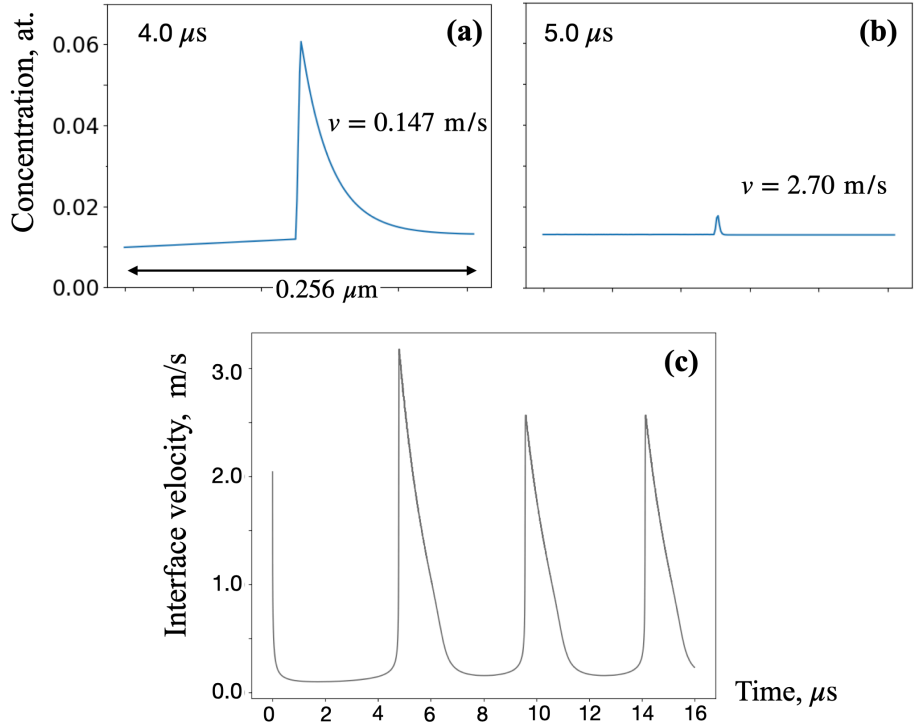

FIG.10 Oscillatory growth under directional cooling condition ($G_T = 10^7$ K/m, $R = 6 \times 10^6$ K/s). The initial temperature at the interface is 905K. (a) diffusion-controlled slow interface under the equilibrium condition; (b) attachment-controlled fast interface under the non-equilibrium condition; (c) Interface velocity shows significant oscillation during the interface propagation under a moving temperature gradient.

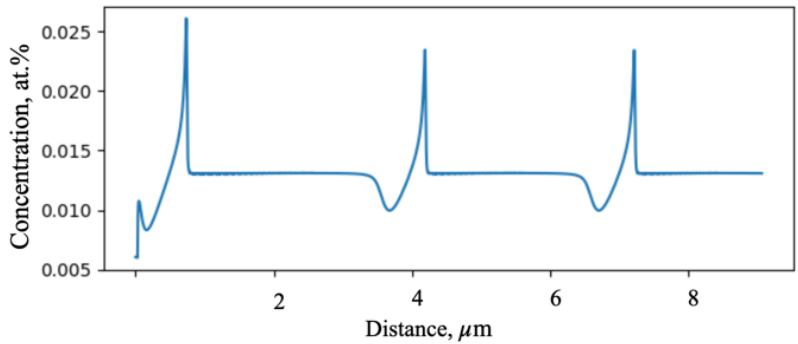

FIG.11 Concentration profile after the pulled growth in Fig.10. The last two spikes have the same heights, suggesting the steady state of the oscillatory interface kinetics.

## V. SUMMARY

A new type of phase-field model that can effectively model alloy solidification under both equilibrium and non-equilibrium conditions is explained and demonstrated. Under equilibrium conditions, the convergence behavior of the dendrite tip is invulnerable to the width effect of the diffuse interface. An acceptable convergence happens when the ratio of diffusion length $l_D$ to interface width $\eta$ approaches 3.5, which is obviously smaller than those in previous quantitative models [11, 13]. Under non-equilibrium conditions, the velocity-dependent partition coefficient from the CG model can be simply input into the PF model. The output one-dimension results agree with the input function. The simulation results are weakly affected even when the $l_D/\eta$ ratio is less than 1. The middle obstacle approach has been proven to be effective in "sharpening" the diffuse interface in phase-field modeling. Future work will relate the capture coefficient to the partition coefficient so that solute diffusion in the solid is included. The velocity-dependent liquidus slope and the effect of solute drag on the interface undercooling are also required to achieve a more accurate simulation. 2D and 3D simulations are to be conducted and compared with analytical theories and experimental observations.

## ACKNOWLEDGMENTS

This work was supported by the Kakenhi Grant-in-Aid for Scientific Research (No. 22J11558) from the Japan Society for Promotion of Science (JSPS).

## REFERENCES

1) W.Kurz and D.J. Fisher, Trans Tech Publications, 1986.
2) J.A. Dantzig and M.Rappaz, EPFL press, second edition, 2016.
3) M.J. Aziz, J. Appl. Phys. **53** 1158-1168 (1982).
4) M. J. Aziz and W. J. Boettinger, Acta Metall. **42**, 527-537 (1994): .
5) S. L. Sobolev, Phys. Rev. E **55**, 6845 (1997).
6) Y.Yang, H.Humadi, D. Buta, B. B. Laird, D. Sun, J. J. Hoyt and M. Asta, Phys. Rev. Let. **107**, 025505 (2011).
7) N. A. Ahmad, A. A. Wheeler, William J. Boettinger and Geoffrey B. McFadden, Phys. Rev. E **58**, 3436 (1998).
8) S. G. Kim, W. T. Kim and T. Suzuki, Phys. Rev. E **60**, 7186 (1999).
9) D. Danilov and B. Nestler. Acta Mater. **54**, 4659-4664 (2006).
10) P. K. Galenko, E. V. Abramova, David Jou, D. A. Danilov, V. G. Lebedev and D. M. Herlach. Phys. Rev. E **84**, 041143 (2011).
11) A. Karma, Phys. Rev, Let. **87**, 115701 (2001)
12) B. Echebarria, R. Folch, A. Karma and M. Plapp. Phys. Rev. E **70**, 061604 (2004).
13) M. Ohno and M. Kiyotaka. Phys. Rev. E **79**, 031603 (2009).
14) A. Carré, B. Böttger and M. Apel. J. Cryst. Growth **380**, 5-13 (2013).
15) P. Tatu and N. Provatas. Acta Mater. **168**, 167-177 (2019).
16) S. Kavousi and A. Z. Mohsen, Acta Mater. **205**, 116562 (2021).
17) K. Ji, D. Elaheh, J. C. Amy and A. Karma. Phys. Rev. Let. **130**, 026203 (2023).
18) I. Steinbach, L. Zhang and M. Plapp, Acta Mater. **60**, 2689-2701 (2012).
19) L. Zhang, E. V. Danilova, I. Steinbach, D. Medvedev and P. K. Galenko. Acta Mater. **61**, 4155-4168 (2013).
20) A. Mukherjee, J. A. Warren and P. W. Voorhees. A quantitative variational phase field framework. Acta Mater. 251, 118897 (2023).
21) W. J. Boettinger, J. A. Warren, C. Beckermann, and A. Karma, Annu. Rev. Mater. Res. **32**, 163-194 (2002).
22) C. Beckermann, H. J. Diepers, I. Steinbach, A. Karma and X. Tong, J. Comput. Phys. **154**, 468-496 (1999).
23) I. Steinbach, Modell. Simul. Mater. Sci. Eng. **17**, 073001 (2009).
24) R. Trivedi and W. Kurz. Acta Metall. Mater. **42**,15-23 (1994).
25) R. Trivedi and W. Kurz. Int. Mater. Rev. **39**,49-74 (1994).
26) K. M. Beatty and K. A. Jackson. J. Cryst. Growth **211**, 13-17 (2000).
27) Gremaud, M., M. Carrard, and W. Kurz. Acta metallurgica et materialia **39**, 1431-1443 (1991).
28) J. T.McKeown, K. Zweiacker and C. Liu, D. R. Coughlin, A. J. Clarke, J. K. Baldwin, J. W. Gibbs et al. Jom 68 985-999 (2016).

**APPENDIX: Explanation of the collect-and-cast operation and invariant concentration in 2D**

The key feature of the traveling wave solution, which is related to the double obstacle free energy density, is that there are singularities at the boundary between the bulk region and interface region. This is quite different from the phase field model with a double well free energy density in its free energy functional. The choice of traveling wave solution has many benefits [23]. In the current model, the reason for choosing this solution is that it enables zeros fluxes at the interface limits (Sec. 2.2) and facilitates the collect-cast operation, which is adopted to avoid undesired solute trapping in the diffuse interface.

The collect and cast operation might easily be understood in 1D, as shown in Fig.4. Here, it is illustrated in the 2D domain to clarify the detailed process. In Fig. A1, the distribution of phase field in the numerical grids is denoted by the grey scale of the dots. The polygonal lines that connect the dots are back, middle and front obstacles from the inner to the outer sides. After the evolution of the phase field, the forward motion of the interface comes from the increment of the phase field near and in the diffuse interface. The amount of rejected solute due to the local transformation of liquid to solid can then be determined according to Eq.(12). As the solid-line arrows denote, the rejected solute is cast just ahead of the middle obstacle following a path normal to the interface. Similarly, the dashed-line arrows denote the path of searching invariant concentration just behind the middle obstacle. The path of these arrows is calculated by the following steps:

**Step 1**. Calculate the interface normal vector at the initial grid $(i, j)$, which is

$\vec{n} = (n_x, n_y)$.

**Step 2**. Search the next grid by referring to the interface normal. Because both $n_x$ and $n_y$ are smaller than 1, the next grid should be determined by rounding the above indexes to be 0 or 1. The grid indexes are

$(i + [n_x], j + [n_y])$.

Accordingly, the next next grid indexes are

$(i + [2n_x], j + [2n_y])$.

The grid search should repeat until the phase field is lower than 0.5.

**Step 3**. The rejected solute should be subtracted from the initial grid and added to the grid just ahead of the middle obstacle.

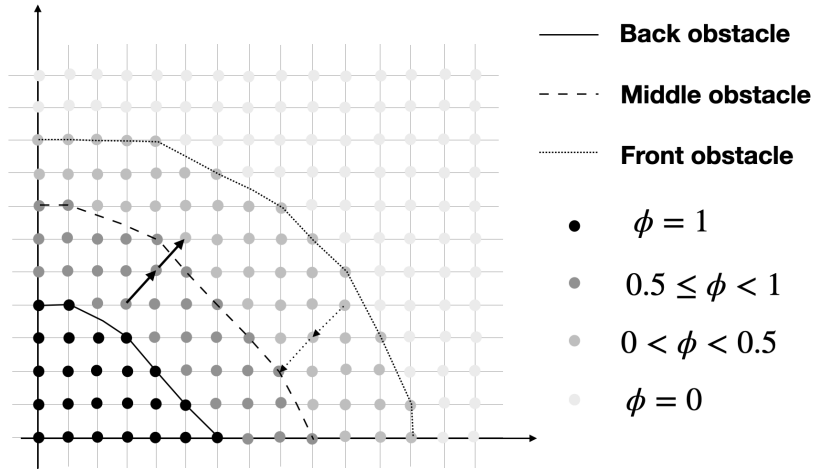

**FIG. A1**. Schematic image of the grid-by-grid distribution of phase field in two dimensions. The solid-line arrows show the path of casting the rejected solute. The dashed-line arrows show the path of searching invariant concentration.